\documentclass[reprint,
superscriptaddress,
 amsmath,
 aps,
prd,
]{revtex4-2}

\usepackage{soul}
\usepackage{graphicx}
\usepackage{dcolumn}
\usepackage{bm}
\usepackage{multirow}
\usepackage{amssymb}
\usepackage{pifont}
\usepackage[colorlinks]{hyperref}
\usepackage{booktabs}

\usepackage[table]{xcolor}

\begin{document}

\preprint{APS/123-QED}

\title{On the Destabilization of High-Mass Neutron Stars by the Emergence of $d^*$-Hexaquarks}
\author{Marcos O. Celi} \email{mceli@fcaglp.unlp.edu.ar}
\affiliation{Grupo de Astrof\'isica de Remanentes Compactos,\\ Facultad de Ciencias Astron{\'o}micas y
  Geof{\'i}sicas, Universidad Nacional de La Plata,\\ Paseo del Bosque
  S/N, La Plata (1900), Argentina.}
\affiliation{CONICET, Godoy Cruz 2290, Buenos Aires (1425), Argentina.}

\author{Mikhail Bashkanov} \email{mikhail.bashkanov@york.ac.uk}
\affiliation{Department of Physics, University of York, , Heslington, York, Y010 5DD, UK}

\author{Mauro Mariani} \email{mmariani@fcaglp.unlp.edu.ar}
\affiliation{Grupo de Astrof\'isica de Remanentes Compactos,\\ Facultad de Ciencias Astron{\'o}micas y
  Geof{\'i}sicas, Universidad Nacional de La Plata,\\ Paseo del Bosque
  S/N, La Plata (1900), Argentina.}
\affiliation{CONICET, Godoy Cruz 2290, Buenos Aires (1425), Argentina.}

\author{Milva G. Orsaria} \email{morsaria@fcaglp.unlp.edu.ar}
\affiliation{Grupo de Astrof\'isica de Remanentes Compactos,\\ Facultad de Ciencias Astron{\'o}micas y
  Geof{\'i}sicas, Universidad Nacional de La Plata,\\ Paseo del Bosque
  S/N, La Plata (1900), Argentina.}
\affiliation{CONICET, Godoy Cruz 2290, Buenos Aires (1425), Argentina.}
\affiliation{Department of Physics, San Diego State University, 5500
  Campanile Drive, San Diego, CA 92182, USA.}
  
\author{Alessandro Pastore} \email{alessandro.pastore@cea.fr}
\affiliation{CEA, DES, IRESNE, DER, SPRC, F-13108 Saint Paul Lez Durance, France}

\author{Ignacio F. Ranea-Sandoval} \email{iranea@fcaglp.unlp.edu.ar}
\affiliation{Grupo de Astrof\'isica de Remanentes Compactos,\\ Facultad de Ciencias Astron{\'o}micas y
  Geof{\'i}sicas, Universidad Nacional de La Plata,\\ Paseo del Bosque
  S/N, La Plata (1900), Argentina.}
\affiliation{CONICET, Godoy Cruz 2290, Buenos Aires (1425), Argentina.}

\author{Fridolin Weber} \email{fweber@sdsu.edu}
\affiliation{Department of Physics, San Diego State University, 5500
  Campanile Drive, San Diego, CA 92182, USA.}  \affiliation{Center for
  Astrophysics and Space Sciences, University of California,\\ San
  Diego, La Jolla, CA 92093, USA.}

\begin{abstract}
We study the effects of the first nontrivial hexaquark, $d^*$(2380), on the equation of state of dense neutron star matter and investigate the consequences of its existence for neutron stars. The matter in the core regions of neutron stars is described using density-dependent relativistic mean-field theory. Our results show that within the parameter spaces examined in our paper, (i) the critical density at which the $d^*$ condensate emerges lies
  between 4 and 5 times the nuclear saturation density, (ii)  $d^*$ hexaquarks are found to exist only
  in rather massive neutron stars, (iii) only relatively small fractions of the matter in the core of a massive neutron star may contain hexaquarks.
\end{abstract}

\maketitle


\section{\label{sec:level1}Introduction}

The advent of new powerful instruments like the LIGO/Virgo gravitational wave detectors and NICER X-ray telescope has revolutionized the approach to studying Neutron Stars (NS) and presents new challenges in comprehending the ultra-dense nuclear matter that constitutes them. The analysis of LIGO/Virgo data from the binary neutron star merger events GW170817 \cite{Abbot:2017gwa} and GW190425 \cite{Abbott:2020goo} in addition to the observations of $2~M_{\odot}$ binary pulsars such as PSR J1614-2230 \cite{Demorest2010}, PSR J0348+0432 \cite{antoniadis2013}, PSR J2215+5135 \cite{Linares2018}, PSR J0740+6620 \cite{Cromartie2020}, combined with NICER data of PSR J0030+0451 \cite{Riley2019anv, Raaijmakers_2019, Miller2019, Bilous2019} have set strong constraints on the theoretical models of the Equation of State (EoS) for matter at densities higher than nuclear saturation density, $n_0$.

In particular, for the isolated pulsar PSR J0030+0451, there are two independent measurements of its gravitational mass and equatorial radius obtained by the NICER collaboration: $M=1.34^{+0.15}_{-0.14} M_\odot$,  $R_{\rm{eq}}=12.71^{+1.14}_{-1.19} $ km \cite{Riley2019anv} and $M=1.44^{+0.15}_{-0.14} M_\odot$,  $R_{\rm{eq}}=13.02^{+1.24}_{-1.06}$ km \cite{Miller2019}. Using data from NICER together with observations of XMM-Newton, two independent mass and radius estimates of PSR J0740+6620 have been inferred \citep{riley2021ApJ-j0740,miller2021ApJ-j0740} (see also the estimates of mass and radius obtained after the revised analysis of the best available data from NICER \citep{salmi:2022ApJ}). The results show that despite being almost 50$\%$ more massive than PRS J0030+0451, the radius of PSR J0740+6620 does not exhibit a significantly smaller value. These observations challenge our understanding of matter inside the inner cores of NS and discard some modern EoS. On the other hand, there is a constraint on the radius of a $1.4 \, M_\odot$ NS \cite{Lattimer2012},
which is determined to be $R_{1.4}<13.6$~km. Additionally, it is estimated that a NS cannot sustain a mass exceeding $\sim 2.3 M_{\odot}$ \cite{Shibata2019}. 
The latter restriction implies that the radius of a NS with a mass of $1.4\, M_\odot$ is estimated to be \mbox{$R_{1.4}=11.0^{+0.9}_{-0.6}$ km.} Moreover, observations based on X-ray emissions, as presented in the study by \citet{Landry2020}, suggest a radius of $R_{1.4}=12.32^{+1.09}_{-1.47}$ km for a NS with the same mass.

In addition to the aforementioned observations, there are also data available from the (presumably) second binary NS merger event GW190425 \cite{Abbott:2020goo}.
In this case, no electromagnetic counterpart has been detected. 
 To estimate the mass of the primary object, two scenarios were considered: one assuming a low spin and the other assuming a high spin. The estimated values for the mass of the primary object in the low-spin scenario range from $1.60$ to $1.87\,M_\odot$, while in the high-spin scenario, the estimated mass ranges from $1.61$ to $2.52\,M_\odot$. The estimated mass values for the secondary object in the low-spin scenario range from $1.49$ to $1.69\,M_\odot$, while in the high-spin scenario, the estimated mass ranges from $1.12$ to $1.68\,M_\odot$. Due to the lack of an electromagnetic counterpart, the constraints on the mass and radius of this NS are not as tight as the ones obtained from GW170817. 
 However, despite this limitation, the detection implies that a NS with a high mass ($M > 1.7 \,M_\odot$) would likely have a larger radius, estimated to be around $R \sim 11$ km or more. 

To further our understanding of NS composition, it is crucial to possess a comprehensive understanding of the nuclear EoS. However, delving into this subject presents significant challenges, primarily due to the inherent complexity of the nuclear many-body problem. Some of the dense-matter studies use a phenomenological approach to describe nuclear interactions and some seek a microscopic many-body perspective. Among the phenomenological approaches, the most frequently employed ones are based on the Skyrme \cite{Skyrme:1959zz, Vautherin1972, Grasso2019} interaction and variations of the relativistic mean-field (RMF) model \cite{walecka:1974ato,Boguta1977,Sedrakian:2023PPNP}. We will focus on the latter in this work. The RMF model used for this study describes interactions between baryons in terms of meson exchanges, based on effective Lagrangian densities \cite{Typel:1999rmf, Typel:2018rmf}. In order to determine the appropriate baryons-mesons coupling constants for these models, we will consider contemporary limitations imposed by nuclear and astrophysical conditions \cite{Pang:2021npm, Huth:2022cns}.

Initially, the cores of NS were believed to consist of a neutron-rich fluid in $\beta$ equilibrium. However, given the exceptionally high densities involved, it is anticipated that new microscopic degrees of freedom will emerge in the cores of NS (for a recent review, see \cite{Sedrakian:2023PPNP} and references therein). It has been found that $\Delta$ particles may constitute a considerable fraction of total particles in NS matter when the density is a few times $n_0$ \cite{KATAYAMA2015,Fortin2017,Sedrakian:2023PPNP}.
It has also been shown that the presence of $\Delta$ particles has a significant impact on the NS properties. More precisely, the $\Delta$ population affects the radii of NS \cite{Schurhoff2010,Cai2015,Zhu2016,Malfatti:2020dba,Sedrakian:2023PPNP}. On top of that, particles containing strange quarks may also be expected to appear \cite{weber:2001sin,WEBER2005193,tolos:2020sin}. Hyperons may be present in the NS interiors and their presence has an appreciable impact on the radii of NS and also on their maximum masses \cite{bednarek:2012hin,fortin:2015nsw}. Another theoretical possibility that has been explored is the appearance of quark matter in the inner cores of compact stars (see, for example, Refs. \cite{alford:2004dqm,orsaria:2019pti,baym:2018fht} and references therein). Such matter could potentially be formed either by free quarks or by quarks forming a color superconducting state
(see, for example, Ref. \cite{alford:2003csw}).

Furthermore, previous studies have examined the significance of the $d^*$(2380) dibaryon
for the nuclear equations of state (EoS) (e.g., \citet{Bashkanov:2019epo}, \citet{Vidana2018}, \citet{Mantziris:2020nsm}). The $d^*$(2380) is the first known non-trivial hexaquark for which experimental evidence is available \cite{Adlarson2011}. It is a massive, positively charged non-strange particle with an integer spin ($J=3$). The study conducted by \citet{bashkanov2019} has revealed that, despite its substantial mass, the $d^*$(2380) dibaryon may be expected to exist at densities at which $\Delta$ particles or hyperons exist in the interiors of NS.

The $d^*$(2380) has very large $\Delta\Delta$ coupling, so that in some theoretical models it is treated as a 70~MeV bound state of two-$\Delta$'s, a Deltaron \cite{Bashkanov:2019epo}. That is why the presence of the $d^*$(2380) in an EoS model substantially changes the behavior of other baryons, substituting $\Delta$'s where possible. Another important point is $d^*$(2380)-mediated many-body forces. The $d^*$(2380) particle, due to the $p n\to d^* \to p n$ process, introduces an additional degree of freedom in nucleon-nucleon interactions. As a result, the $d^*-N$ interaction effectively plays the role of three-body (3N) nucleon forces, while $d^*-d^*$ interactions resemble four-body (4N) forces, which might be important in high-density matter. While our Lagrangian does not explicitly incorporate these interactions, the mere presence of the $d^*$(2380) particle partially accommodates such dynamics. 

The main objective of this work is to build upon the analysis of the 
behavior of the $d^*$(2380), as presented in \citet{Mantziris:2020nsm}. 
We aim to enhance the aforementioned study by incorporating a broader set of EoS, introducing density-dependent coupling constants, and ensuring that the resulting EoS comply with both the latest astrophysical 
constraints as well as constraints derived from nuclear theory \cite{Annala2020}. We will incorporate hyperons and the $\Delta$ resonance as potential constituents of dense NS matter. Considering that the work by \citet{Mantziris:2020nsm} highlights the destabilizing effect of $d^*$(2380) particles in NS, our main objective is to examine whether the parameters within the EoS models of our study permit the presence of $d^*$(2380) particles in dense NS matter.
The EoS models we will use in this study are DD2 \cite{Typel:2009a,Typel:2018DD}, GM1L \cite{Spinella2017:thesis,Spinella2018:u}, and SW4L \cite{Spinella:2019hns, Malfatti:2020dba}, which are computed within the framework of density-dependent RMF theory. Our choice of these three parametrizations is motivated by their ability to satisfy the constraint of 2 $M_{\odot}$ for NSs, which is a critical benchmark if you consider hyperons, and their capacity to introduce improvements in RMF models with fixed coupling constants. We have chosen these three parametrizations as representatives of a wide family of hadronic EoS compatible with modern astronomical observations. In these models, the coupling constants associated with hyperons exhibit density dependence, while for the $d^*$(2380) particle, we will consider the coupling constant to be density-independent. Given the lack of understanding of the d$^{\ast}$ couplings within the nucleonic sector,  including density-dependent coupling constants for the d$^{\ast}$ would necessitate additional free parameters, increasing the number of free parameters further. We note, however, that our framework has the capability to be extended to a density-dependent version of the coupling constant for the $d^*$(2380), which will be the topic of a future study. We also note that upcoming ground-based experiments focusing on $d^*$(2380) photoproduction in nuclei \cite{Bashkanov:2020sot} hold the potential to provide valuable insight into the coupling constant of the $d^*$(2380) by measuring the nuclear-dependent medium modification of the $d^*$(2380) mass.

The paper is structured in the following manner. In Section \ref{sec:RMFd*}, 
we provide a detailed description of the density-dependent RMF model used in this study and explain how the inclusion of $d^*$(2380) is incorporated within this theoretical framework. In Section \ref{sec:results}, we discuss the astrophysical implications of the appearance of the $d^*$(2380) in the cores of NS and provide an analysis of the constraints imposed by current astronomical observations on the coupling constants associated with this particle. In Section \ref{sec:summary}, we summarize the main findings and present the most relevant conclusions drawn from our research.

\section{The RMF model considering $d^*$(2380)} \label{sec:RMFd*}

The Lagrangian density has a general form that includes various components: baryons (nuclear matter $N=n,p$), the four states of the $\Delta$ particle ($\Delta=\Delta^-,\Delta^0,\Delta^+,\Delta^{++}$), hyperons \mbox{($H=\Lambda^0, \Sigma^+, \Sigma^0, \Sigma^-, \Xi^0, \Xi^-, \Omega^-$),} mesons $\sigma$, $\omega$, $\rho$, $\sigma^*$\footnote{In particle physics nomenclature this particle is usually referred as $f_0(980)$.}, $\phi$, non-linear terms for the $\sigma$ meson, the dibaryon $d^*$(2380), and leptons ($l= e^-, \mu^-$). It is given by \cite{Malfatti:2020dba}
\begin{equation}\label{eq:lagrangian}
\mathcal{L}=\sum_B \mathcal{L}_B+\mathcal{L}_{\sigma\omega\rho}+\mathcal{L}_{NL\sigma}+\mathcal{L}_{\phi\sigma^*}+\mathcal{L}_{d^*}+\sum_l \mathcal{L}_l,
\end{equation}
where
\begin{eqnarray}
  \mathcal{L}_B &=& \sum\limits_B \overline\psi_B\bigl[\gamma_{\mu}
    (i\partial^{\mu}-g_{\omega B}\omega^{\mu}-g_{\phi
      B}\phi^{\mu}-\tfrac{1}{2}g_{\rho B}\boldsymbol{\tau}\cdot
    \boldsymbol{\rho}^{\mu})\nonumber\\ &&-(m_B-g_{\sigma
      B}\sigma-g_{\sigma^*
      B}\sigma^*)\bigr]\psi_B.
      \label{eq:Lag_bar}
\end{eqnarray}
The summation over $B$ includes all baryons including the $\Delta$ resonance. The mesonic Lagrangians are as follows, 
\begin{eqnarray}
\mathcal{L}_{\sigma\omega\rho}&=&\tfrac{1}{2}\left(\partial_{\mu}\sigma\partial^{\mu}\sigma-m^2_{\sigma}\sigma^2\right)-\tfrac{1}{4}\omega_{\mu\nu}\omega^{\mu\nu}+\tfrac{1}{2}m^2_{\omega}\omega_{\mu}\omega^{\mu}\nonumber\\&&
  -\tfrac{1}{4}\boldsymbol{\rho}_{\mu\nu}\cdot\boldsymbol{\rho}^{\mu\nu}
  +\tfrac{1}{2}m^2_{\rho}\boldsymbol{\rho}_{\mu}\cdot\boldsymbol{\rho}^{\mu}, 
\end{eqnarray}
\begin{eqnarray}
\mathcal{L}_{NL\sigma}&=&-\tfrac{1}{3}\tilde{b}_{\sigma}m_n\left(g_{\sigma
    N}\sigma\right)^3-\tfrac{1}{4}\tilde{c}_{\sigma}\left(g_{\sigma
    N}\sigma\right)^4 , 
    \label{eq:NLsigma} 
\end{eqnarray}
\begin{eqnarray}
  \mathcal{L}_{\phi\sigma^*} &=& -\tfrac{1}{4}\phi^{\mu\nu}\phi_{\mu\nu}+\tfrac{1}{2}m^2_{\phi}\phi_{\mu}
  \phi^{\mu}\nonumber\\&&+\tfrac{1}{2}\left(\partial_{\mu}
  \sigma^*\partial^{\mu}\sigma^* -m^2_{\sigma^*}\sigma^{*2}\right) , 
\end{eqnarray}
The dibaryon Lagrangian reads  \cite{Mantziris:2020nsm}
\begin{eqnarray}
\mathcal{L}_{d^*}=\mathcal{D}^*\xi_{d^*}^* \mathcal{D} \xi_{d^*}-m^{* 2}_{d^*} \xi_{d^*}^* \xi_{d^*} , 
\end{eqnarray}
where $\mathcal{D}=\left(\partial_\mu+i g_{\omega d^*} \omega_\mu\right)$, $m^*_{d^*}=m_{d^*}-g_{\sigma d^*} \sigma$, and $\xi_{d^*}$ is the dibaryon isoscalar-scalar field. Note that this description adopted in Ref.~\cite{Mantziris:2020nsm} from Ref.~\cite{Faessler:1998din} corresponds to a spin $S=0$ particle and it neglects spins structure of the $d^*$ hexaquark which is $S=3$ particle. A proper spin $S=3$ theoretical treatment can be found in Ref.~\cite{Berends:1980os3}. Following Ref.\cite{Mantziris:2020nsm} we used a simplified $d^*$ treatment. We do not expect that such ommissions would massively change the outcome of our calculations.

The leptons are described by
\begin{eqnarray}
  \mathcal{L}_l =\bar{\Psi}_l\left(i \gamma_\mu \partial^\mu-m_l\right) \Psi_l , 
\end{eqnarray}
Baryon-baryon interactions are modelled in terms of scalar ($\sigma, ~\sigma^*$), vector ($\omega,~ \phi$), and isovector
($\rho$) meson fields. 

The parametrization of the density-dependent constants accounting for nuclear medium effects are given by \cite{particles2017, Fuchs1995}
\begin{equation}
g_{i B}(n)=g_{i B}\left(n_0\right) a_i \frac{1+b_i\left(\frac{n}{n_0}+d_i\right)^2}{1+c_i\left(\frac{n}{n_0}+d_i\right)^2}\,,
\label{eq:so_func}
\end{equation}
for $i=\sigma,\omega$, and
\begin{equation}
g_{\rho B}(n)=g_{\rho B}\left(n_0\right) \exp \left[-a_\rho\left(\frac{n}{n_0}-1\right)\right]\,.
\label{eq:rho_func}
\end{equation}
The constants $a_i$, $b_i$, $c_i$ and $d_i$ are fixed, in the nucleonic sector, by the binding energies, charge and diffraction radii, spin-orbit splitting, and the neutron skin thickness of finite nuclei (see Table \ref{table:parametrizations} in Section \ref{sec:App} for details). The density dependence of the meson-baryon coupling makes it unnecessary to add self-interactions of the $\sigma$ meson in the DD2 parametrization ($\mathcal{L}_{NL\sigma}$=0). Hence, only the GM1L and SW4L  parametrizations need the nonlinear terms shown in Eq.~\eqref{eq:NLsigma} (see also Table \ref{table:models}). 
As the $g_{\omega d^*}$ coupling is responsible for repulsion and $g_{\sigma d^*}$ for attraction (following the sign convention of Ref.~\cite{Mantziris:2020nsm}), we limit our analysis to negative values of $g_{\omega d^*}$ and positive values of $g_{\sigma d^*}$.
The range of the $d^*$(2380) coupling constant was chosen between $g_{i d^*}=0$ 
(representing the non-interacting case) and $g_{i d^*}=2\, g_{iN}$ (corresponding to the universal coupling limit)\footnote{There are three proposed symmetry-inspired possibilities: 1) the $d^*$ do not interact with nuclear matter $x_{\sigma d^*}=x_{\omega d^*}=0$; 2) the $d^*$ interacts with nuclear matter in exactly the same way like nucleons $x_{\sigma d^*}=-1\cdot x_{\omega d^*}=1$; 2) $x_{\sigma d^*}=-1\cdot x_{\omega d^*}=2$ - quark counting unified scheme.}. The meson-hyperon coupling constants of all parametrizations have been determined following the Nijmegen extended soft core (ESC08) model (see \cite{Spinella:2019hns, Malfatti:2019hqm, Malfatti:2020dba}, and references therein). Details of the determination of some of these couplings are given in Section \ref{sec:App}. For the $\Delta$-resonance we use a quasi-universal meson--$\Delta$ coupling \mbox{$x_{{\sigma} {\Delta}} = x_{{\omega} {\Delta}} = 1.05$},   $x_{{\rho}
  {\Delta}} =x_{{\phi} {\Delta}} =1.0$, $x_{{\sigma^*} {\Delta}} =0.0\,$. Following \citet{Malfatti:2020dba} we  also use  \mbox{$x_{{\sigma} {\Delta}} = x_{{\omega} {\Delta}} = 1.25$.} At this point it is important to mention that although $\Delta$-resonances are spin 3/2 particles, their equation of motion can be written as those of spin 1/2 and for this reason can be included in the Lagrangian of Eq.~\eqref{eq:Lag_bar} (see Ref.~\cite{DePaoli:2013rsp} for details). For an easier numerical treatment we consider \mbox{$x_{ i H} = g_{i H}/g_{i N}$} for all meson--hadron coupling constants.

\begin{table}[]
\begin{tabular}{cccc}
\toprule
& ~~~DD2~~~ & ~~~GM1L~~~ & ~~~SW4L~~~ \\
\midrule
$g_{\sigma^* B}$ & \ding{55} & \ding{55} & \ding{51} \\
$g_{\phi B}$ & \ding{55} & \ding{55} & \ding{51} \\
$\mathcal{L}_{\phi \sigma^*}$ & \ding{55} & \ding{55} & \ding{51} \\
$\mathcal{L}_{NL\sigma}$ & \ding{55} & \ding{51} & \ding{51} \\
$g_{i B}(n)$ & \ding{51} & \ding{55} & \ding{55} \\
$g_{\rho B}(n)$ & \ding{51} & \ding{51} & \ding{51} \\
\bottomrule
\end{tabular}
\caption{Coupling constants and terms included (indicated with a \ding{51}) or excluded (\ding{55}) from the general Lagrangian of Eq.~\eqref{eq:lagrangian} depending on the hadronic parametrizations DD2, GML1 \cite{Spinella:2018nei, Spinella;2020qmi} (and references therein), and SW4L \citep{Spinella:2019hns, Malfatti:2020dba}.}
\label{table:models}
\end{table}

\begin{table*}[htb]
\begin{tabular}{cccc}
\toprule 
Saturation Properties & ~~~DD2 \cite{Typel:2009a,Typel:2018DD}~~~ 
& ~~~GM1L \cite{Spinella2017:thesis,Spinella2018:u}~~~ & ~~~SW4L \cite{Spinella:2019hns, Malfatti:2020dba}~~~ \\
\midrule
$n_0$  (fm$^{-3}$)    & 0.149   & 0.153 & 0.150  \\
$E_0$  (MeV)          & $-16.02$  & $-16.30$ & $-16.00$ \\
$K_0$  (MeV)          & 242.7   & 300.0 & 250.0 \\
${m_{N}}^*/m_N$       &  0.56   & 0.70 & 0.70 \\
$J_0$    (MeV)          &  32.8   & 32.5 & 30.3 \\
$L_0$  (MeV)          &  55.3   & 55.0 & 46.5 \\ 
\bottomrule
\end{tabular}
\caption{Properties of nuclear matter at saturation density for the parametrizations used in this work. Shown are the values of the nuclear saturation density $n_0$, energy 
per nucleon $E_0$, nuclear compressibility $K_0$, effective nucleon mass $m_N^*/m_N$, asymmetry energy $J_0$, and the slope of the asymmetry energy $L_0$.}  
\label{table:properties}
\end{table*}
\begin{figure*}[t]
    \centering
    \includegraphics[width=0.85\linewidth]{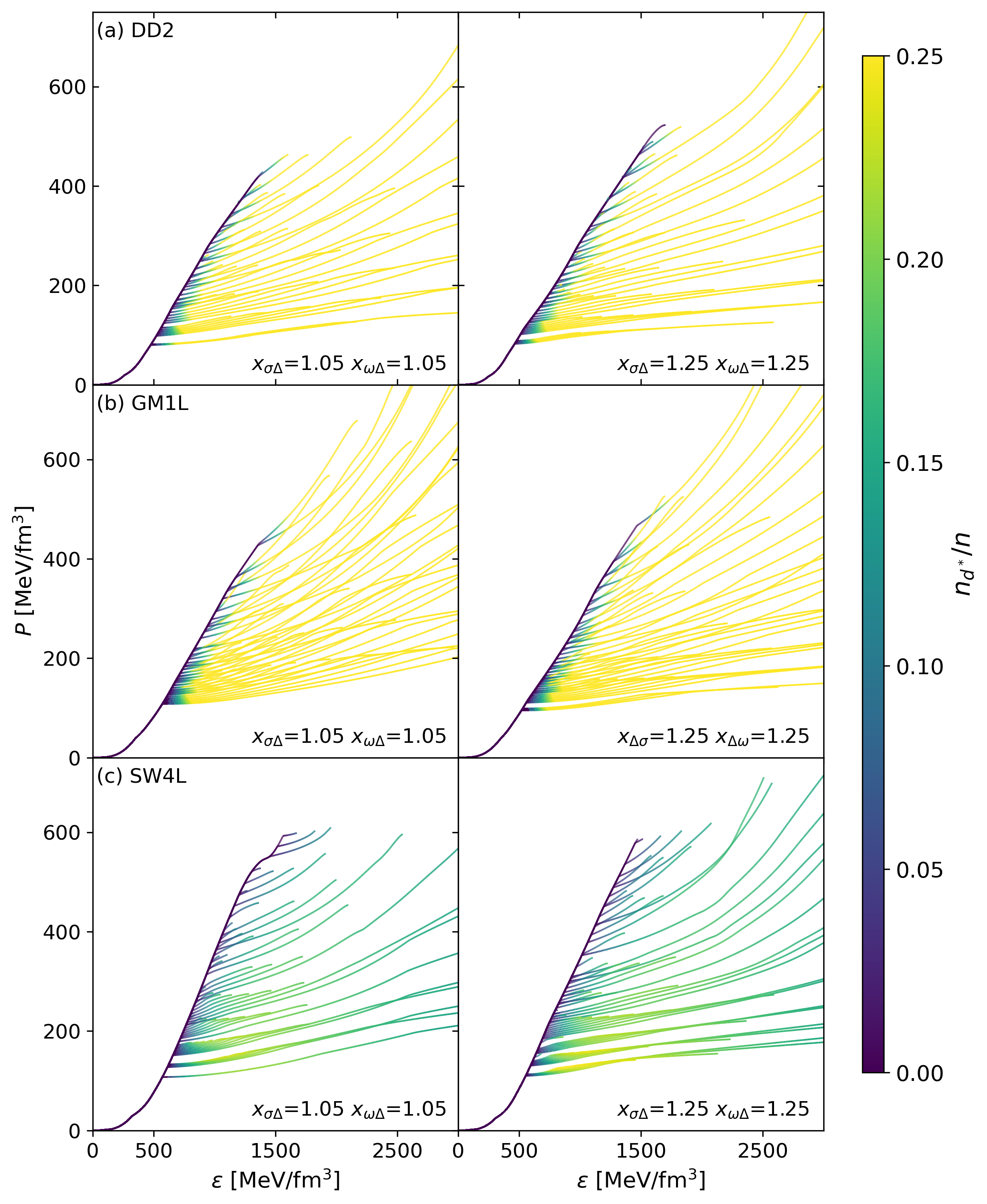}
    \caption{Pressure as a function of the energy density for the DD2 (a), GM1L (b), and SW4L (c) parametrizations, considering $x_{\sigma \Delta} = x_{\omega \Delta} = 1.05$ (left) and $x_{\sigma \Delta} = x_{\omega \Delta} = 1.25$ (right). The color bar represents the ratio of the dibaryon number density to the total baryon number density.}
    \label{fig:eos}
\end{figure*}
The general system of non-linear, coupled equations in the RMF approximation is obtained by deriving the equations of motion that arise from Eq.~\eqref{eq:lagrangian} and subsequently substituting the meson field operators with their mean-field values,
\begin{eqnarray}
m_{\sigma}^2 \bar{\sigma} &=& \sum_{B} g_{\sigma B} n_B^s + \frac{\partial\mathcal{L}_{NL\sigma}}{\partial{\bar{\sigma}}}+ g_{\sigma d^*} n_{d^*} \,, \nonumber
\\ m_{\omega}^2 \bar{\omega}
&=& \sum_{B} g_{\omega B} n_{B} - g_{\omega d^*} n_{d^*}\nonumber ,\label{eq:nonlinear} \\ m_{\rho}^2\bar{\rho} &=&
\sum_{B}g_{\rho B} I_{3B} n_{B} \, , \nonumber
\\ m_{\sigma^*}^2 \bar{\sigma^*} &=&
\sum_{B} g_{\sigma^* B} n_B^s\, , \nonumber
\\ m_{\phi}^2
\bar{\phi} &=& \sum_{B} g_{\phi B} n_{B}\, , 
\end{eqnarray}
where the $d^*$ number density is given by \mbox{$n_{d^*}=2 (m_{d^*} - g_{\sigma d^*} \bar{\sigma}) \xi^*_{d^*}\xi_{d^*}=2 (\mu_{d^*} + g_{\omega d^*} \bar{\omega}) \xi^*_{d^*}\xi_{d^*}$\cite{Mantziris:2020nsm}.} $I_{3B}$ is the 3-component of isospin. The quantities $n_{B}^s$ and $n_{B}$ are the scalar and particle number densities of a baryon, which are given by

\begin{eqnarray}
n_{B}^s&=&{{\frac{2 J_B+1}{2\,\pi^2}}} \int^{p_{F_B}}_0 \,p^2\,dp\,
\frac{m_B^*}{\sqrt{p^2+m_B^{*2}}}, \\ n_{B}&=& {{\frac{2 J_B+1}{6\,\pi^2}}}\,p_{F_B}^3
\, ,
\end{eqnarray}

where $m_B^*= m_B - g_{\sigma B}\bar{\sigma}-g_{\sigma^* B}\bar{\sigma^*}$ denotes the effective baryon mass and $p_{F_B}$ and $J_B$ are, respectively, the Fermi momentum and the total spin of a baryon of type $B$. 
The total baryon number density, $n$, follows from
\begin{equation}
    n = \sum_B n_B .
\end{equation}
The chemical equilibrium condition for the dibaryon is given by \mbox{$\mu_{d^*} = 2 \mu_n - \mu_e$} \cite{Mantziris:2020nsm}, where $\mu_n$ and $\mu_e$ represent the chemical potentials of neutrons and electrons, respectively. In addition, the chemical equilibrium of nucleons, hyperons and $\Delta$-resonances is given by $\mu_B = \mu_n + q_B\,\mu_e$,  where $q_B$ is the corresponding baryon electric charge.
Thus, the chemical potential of a baryon takes the form
\begin{equation}
\mu_B = g_{\omega B} \bar{\omega} + g_{\rho B} \bar{\rho} I_{3B}
+g_{\phi B} \bar{\phi} +\sqrt{p^2_{F_B}+m_B^{*2}}+ \widetilde{R} \, ,
\end{equation}
where the term 
\begin{eqnarray}
\widetilde{R} =\sum_B&&\left( \frac{\partial g_{\omega B}(n)}{\partial
  n} n_B \bar{\omega} + \frac{\partial g_{\rho B}(n)}{\partial n}
I_{3B} n_B \bar{\rho} \right. \nonumber\\ &-& \left.\frac{\partial g_{\sigma
    B}(n)}{\partial n} n_B^s \bar{\sigma}\right) \, , 
\label{eq:rear}
\end{eqnarray}
guarantees thermodynamic consistency \cite{Hofmann:2001aot}.

The hadronic pressure is given by

\begin{eqnarray}
P_H &=& \frac{1}{3}\sum_B \frac{2 J_B+1}{2\,\pi^2} \int^{p_{F_B}}_0 \! dp \, 
\frac{p^4}{\sqrt{p^2+m_B^{*2}}}-\frac{1}{2} m_{\sigma}^2
\bar{\sigma}^2\nonumber\\ &-& \frac{1}{2} m_{\sigma^*}^2
\bar{\sigma^*}^2 + \frac{1}{2} m_{\omega}^2 \bar{\omega}^2 +
\frac{1}{2} m_{\rho}^2 \bar{\rho}^2+ \frac{1}{2} m_{\phi}^2
\bar{\phi}^2\nonumber\\ &-& \frac{1}{3} \tilde{b}_{\sigma} m_N (g_{\sigma N}
\bar{\sigma})^3 - \frac{1}{4} \tilde{c}_{\sigma} (g_{\sigma N}
\bar{\sigma})^4 + n \widetilde{R}\, ,  
\label{eq:pressure}
\end{eqnarray}

and the hadronic energy density can be expressed as

\begin{eqnarray}
\varepsilon_H &=& \sum_B \frac{2 J_B+1}{{{2}}\,\pi^2} \int^{p_{F_B}}_0 \! dp \, 
p^2\, \sqrt{p^2+m_B^{*2}}+\frac{1}{2} m_{\sigma}^2
\bar{\sigma}^2\nonumber\\ &+& \frac{1}{2} m_{\sigma^*}^2
\bar{\sigma^*}^2 + \frac{1}{2} m_{\omega}^2 \bar{\omega}^2 +
\frac{1}{2} m_{\rho}^2 \bar{\rho}^2+ \frac{1}{2} m_{\phi}^2
\bar{\phi}^2\nonumber \\ &-& \mathcal{L}_{NL\sigma} + m_{d^*}^* n_{d^*},  
\label{eq:energy}
\end{eqnarray}

where $m_{d^*}^*=m_{d^*} - g_{\sigma d^*} \bar{\sigma}$ denotes the effective mass of $d^*$
baryons in the RMF approximation. Note that the dibaryon contributes directly to Eq.~\eqref{eq:energy} but also contributes indirectly to Eq.~\eqref{eq:pressure} due to the terms involving the wave function $\xi_{d}^*$ in the system of Eqs.~\eqref{eq:nonlinear}.

It should be noted that for energy densities \mbox{$\varepsilon < 56~\mathrm{MeV~fm^{-3}}$,} corresponding to the crust layers of NS, we use the Baym-Pethick-Sutherland (BPS) and Baym-Bethe-Pethick (BBP) EoS \cite{Baym1971tgs, Baym1971nsm}.

\section{Results} \label{sec:results}

\begin{figure*}
    \centering
    \includegraphics[width=0.8\linewidth]{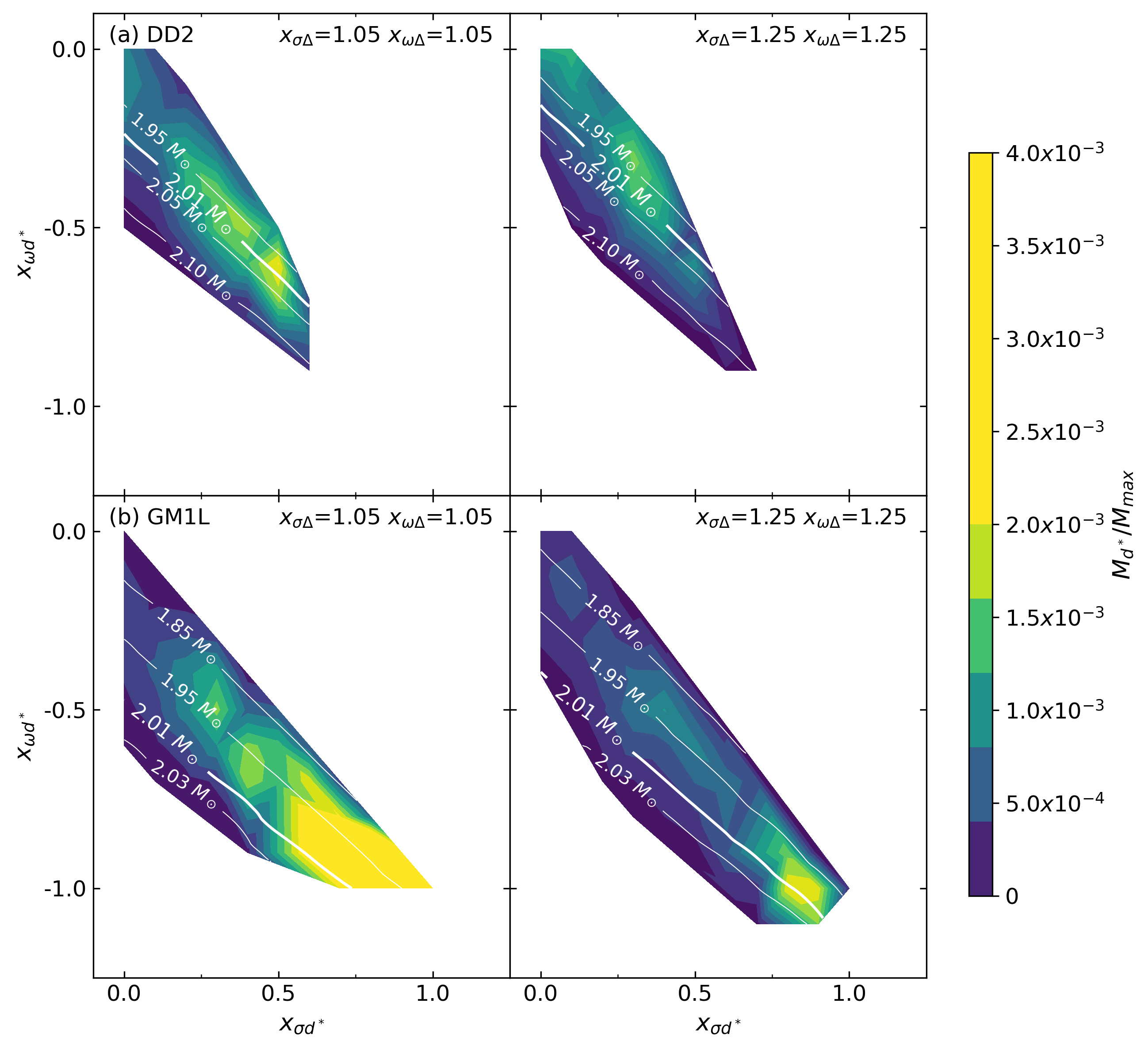}
    \caption{Ratio between the stellar mass containing dibaryons, $M_d^*$, and the maximum stellar mass, $M_{\rm max}$.
This ratio is analyzed in the $x_{\omega d^*}$--$x_{\sigma d^*}$ plane for two different scenarios: one with $x_{\sigma \Delta} = x_{\omega \Delta} = 1.05$ on the left-hand side and the other with $x_{\sigma \Delta} = x_{\omega \Delta} = 1.25$ on the right-hand side. The analysis considers the DD2 and GM1L parametrizations.
      The color bar 
    shows the $M_d^*$ proportion in the maximum-mass stellar configuration. The white lines show the maximum masses as a function of  $x_{\omega d^*}$ and $x_{\sigma d^*}$. 
    (The ratio $M_d^*/M_{\rm max}$ for the SW4L parametrization is below the precision of our calculations and, therefore, is not presented in the figure.)}
    \label{fig:d*_mass}
\end{figure*}

To begin this section, we show  in  panels (a), (b), and (c)  of Fig.~\ref{fig:eos}
the pressure as a function of energy density for the DD2, GM1L, and SW4L parametrizations.
The nuclear saturation properties associated with these EoS are shown in Table \ref{table:properties}.
\begin{figure*}
    \centering
    \includegraphics[width=0.95\linewidth]{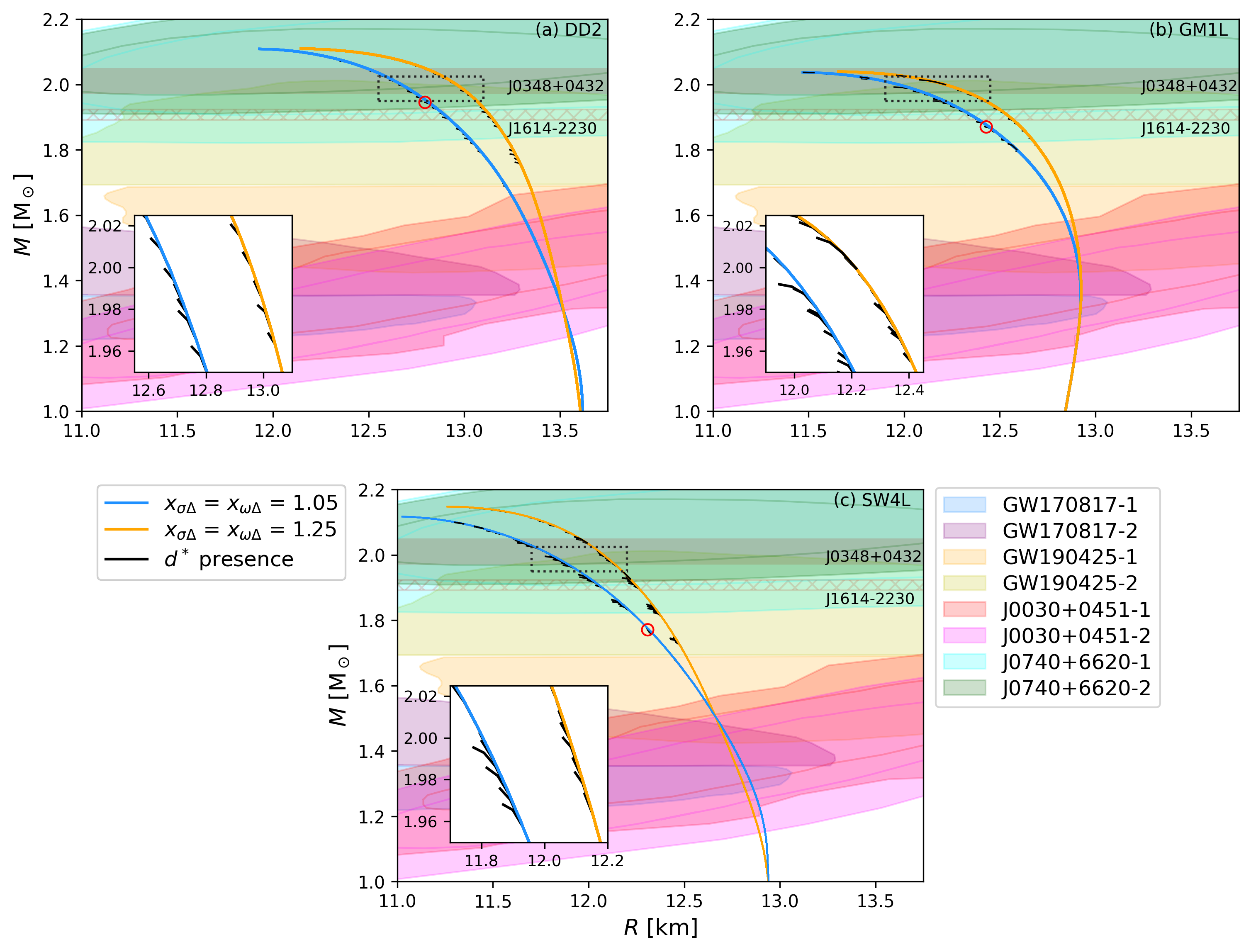}
    \caption{Mass-radius relationship shown for the DD2 (a), GM1L (b), and SW4L (c) parametrizations with different combinations of $x_{\omega d^*}$ and $x_{\sigma d^*}$. The orange (light-blue) curve corresponds to $x_{\sigma \Delta} = x_{\omega \Delta} = 1.25$ ($x_{\sigma \Delta} = x_{\omega \Delta} = 1.05$). The presence of the $d^*$-hexaquark destabilizes NS, resulting in stable, very short branches (black short lines) of stellar 
    configurations containing dibaryons. The enlarged area in each panel shows the effect of the combination of $x_{\omega d^*}$ and $x_{\sigma d^*}$ for stars close to the maximum mass configuration. The red circles indicate the stars considered in Fig.~\ref{fig:profs}.}
    \label{fig:mr}
\end{figure*}
For each parametrization, the left (right) panels display the EoS for meson-$\Delta$ coupling constants \mbox{$x_{\sigma \Delta} = x_{\omega \Delta} = 1.05$} ($x_{\sigma \Delta } = x_{\omega \Delta} = 1.25$). In the exploration of each parametrization, we investigate the parameter space for the ratio of the coupling constants of the  $\sigma$ and $\omega$ mesons, and the $d^*$ dibaryon, with values ranging from $-2 \leq x_{\omega d^*}\leq 0$ and $0 \leq x_{\sigma d^*}\leq 2$.
Each curve in Fig.~\ref{fig:eos} represents a specific combination of $x_{\sigma d^*}$ and $x_{\omega d^*}$, given a different parametrization and a different value of $x_{ \sigma \Delta }$ and $x_{\omega \Delta}$. We find that for a fixed value of $x_{\sigma d^*}$, the smaller (less negative) the absolute value of $x_{\omega d^*}$, the earlier the appearance of the  $d^*$. In addition, for a fixed value of $x_{\omega d^*}$, the larger the absolute value of $x_{\sigma d^*}$, the earlier the $d^*$ particle appears. Furthermore, the appearance of the dibaryon \textit{flattens} the EoS, i.e., the increase of pressure is reduced with increasing energy density. The color bar located to the right of the figure shows the ratio of dibaryons to baryons, $n_{d^*}/n$. It can be seen that after the $d^*$-hexaquark appearance, the ratio of dibaryons increases as the energy density increases. This behaviour is more noticeable for the DD2 and GM1L parametrizations, panels (a) and (b), respectively. Furthermore, the inclusion of hyperon-hyperon interactions, via the strange-scalar ($\sigma^\ast$) and strange-vector ($\phi$) mesons in the SW4L parametrization, results in a stiffening of the hadronic EoS when comparing SW4L with DD2 or GM1L. One can observe the stiffening, setting a specific energy density value, and comparing the pressures for each parametrization. Consequently, the \textit{flattening} of $P(\varepsilon$) in the SW4L parametrization is less pronounced when the $d^{\ast}$ appears. This leads to the absence of the $M_{d^\ast}$ proportion in the configurations with maximum stellar mass for the SW4L parametrization (note that  Fig.~\ref{fig:d*_mass} only shows the panels for DD2 and GM1L parametrizations).

The mass-radius relationship for each parametrization is shown in Fig.~\ref{fig:mr}. The light blue (orange) curve represents a meson-$\Delta$ coupling constant \mbox{$x_{\sigma \Delta}=x_{\omega \Delta} = 1.05~ (x_{ \sigma \Delta }=x_{\omega \Delta}  = 1.25)$.} These colored curves represent configurations without the presence of the $d^*$-hexaquark particle, while the introduction of this particle is indicated by a change to black color on each curve. Consequently, only the black branches indicate stellar configurations with $d^*$ presence. 

As can be seen, the presence of $d^*$-hexaquark particles has a destabilizing effect on NS. If the combination of $x_{\omega d^*}$ and $x_{\sigma d^*}$ leads to the appearance of $d^*$ particles at lower energy densities in the EoS, the corresponding mass-radius curve will be truncated before reaching the maximum-mass configuration. As a result, there are very short stable branches of stellar configurations containing dibaryons. 

Moreover, this behavior is particularly important as it affects the compatibility of the EoS with high mass pulsars such as J1614-2230 \cite{Demorest2010}, J0348+0432 \cite{antoniadis2013}, and J0740+6620 \cite{Cromartie2020}. Stellar configurations with an early occurrence of $d^*$ are unable to explain these high mass pulsars due to the limited stability of dibaryons in their EoS.
The enlarged area in each panel of the figure provides a detailed view of the
influence of the  combination of $x_{\omega d^*}$ and $x_{\sigma d^*}$ for the different  parametrizations. Notably, it is observed that the black stable branches containing dibaryons are slightly longer in panels (b) and (c) for $x_{\omega d^*}=x_{\sigma d^*}= 1.05$ compared to the DD2 parametrization in panel (a).

\begin{figure}[h]
    \includegraphics[width=1.0\linewidth]{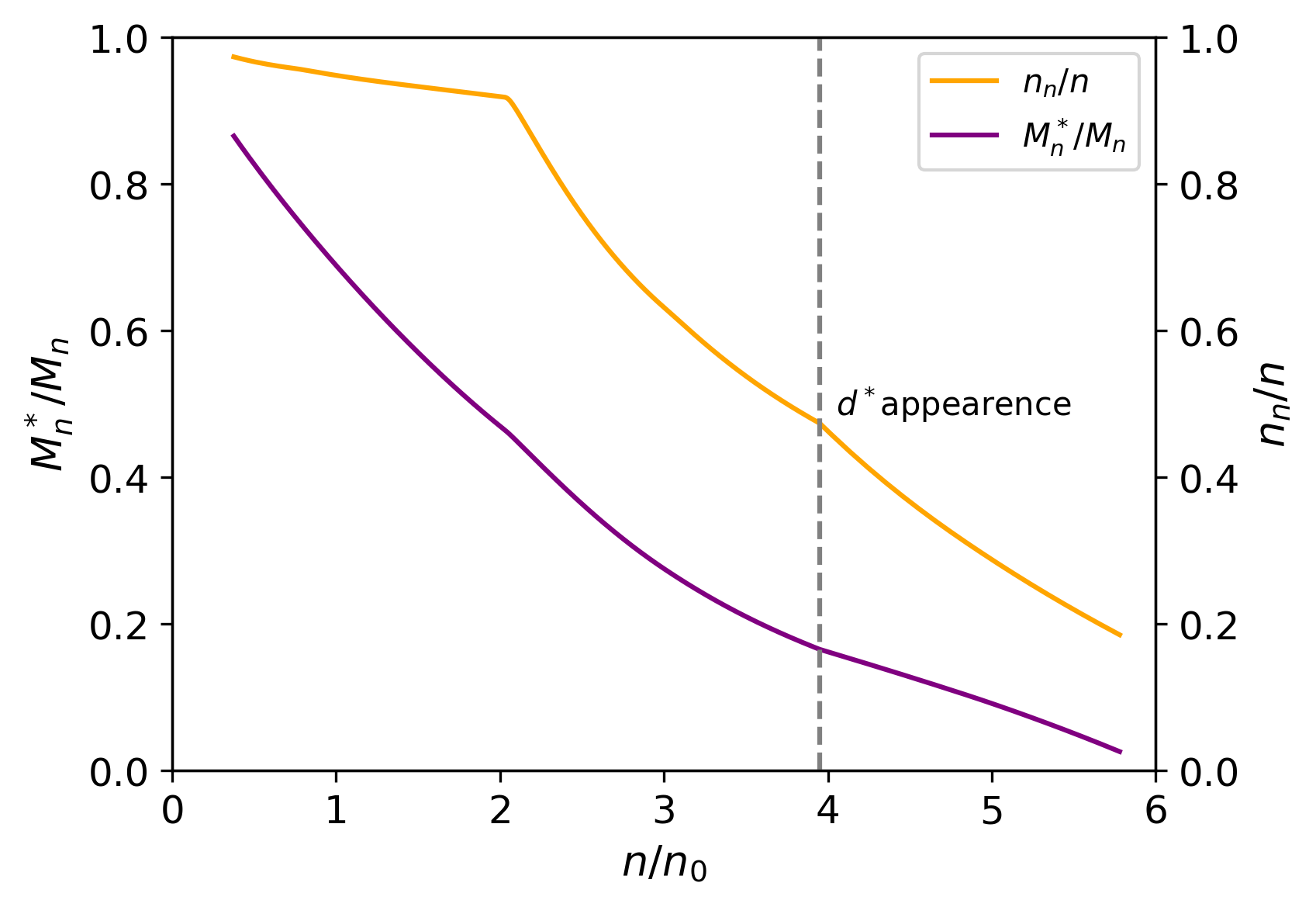}
    \caption{Dynamical neutron mass and the associated number density due to the instabilities discussed in the text for a representative truncated EoS. In particular, we present the results for SW4L parametrization, and $x_{\sigma d^*} = 0.7$ , $x_{\omega d^*}= -0.9$ , $x_{\sigma \Delta} = x_{\omega \Delta} = 1.05$. We show the neutron dynamical mass, in units of the bare neutron mass, $M_n^*/M_n$, and the neutron number density, in units of the baryonic number density, $n_n/n$, as a function of the baryonic number density, in units of $n_0$. The dashed line indicates the onset of the $d^*$. This truncated EoS represents globally the behavior of all truncated EoS in this work.}
    \label{fig:eff_mass}
\end{figure}

For completeness, in Fig.~\ref{fig:eff_mass} we show a representative case of truncated EoS due to the aforementioned instabilities.  We display the neutron dynamical mass, denoted as  $M_n^*/M_n$, in units of the bare neutron mass, and the neutron number density, $n_n/n$, in units of the baryonic number density, as a function of the baryonic number density in units of $n_0$. The onset of the $d^*$ is indicated by a dashed vertical line. The truncation of this EoS is not a consequence of the effective neutron mass reaching zero before achieving the maximum mass configuration; rather, it arises from the effects of different combinations of $d^*$ couplings in each parametrization. It is important to remark that this truncated EoS represents globally the behavior of all truncated EoS in this work.

\begin{figure*}
    \centering
    \includegraphics[width=0.95\linewidth]{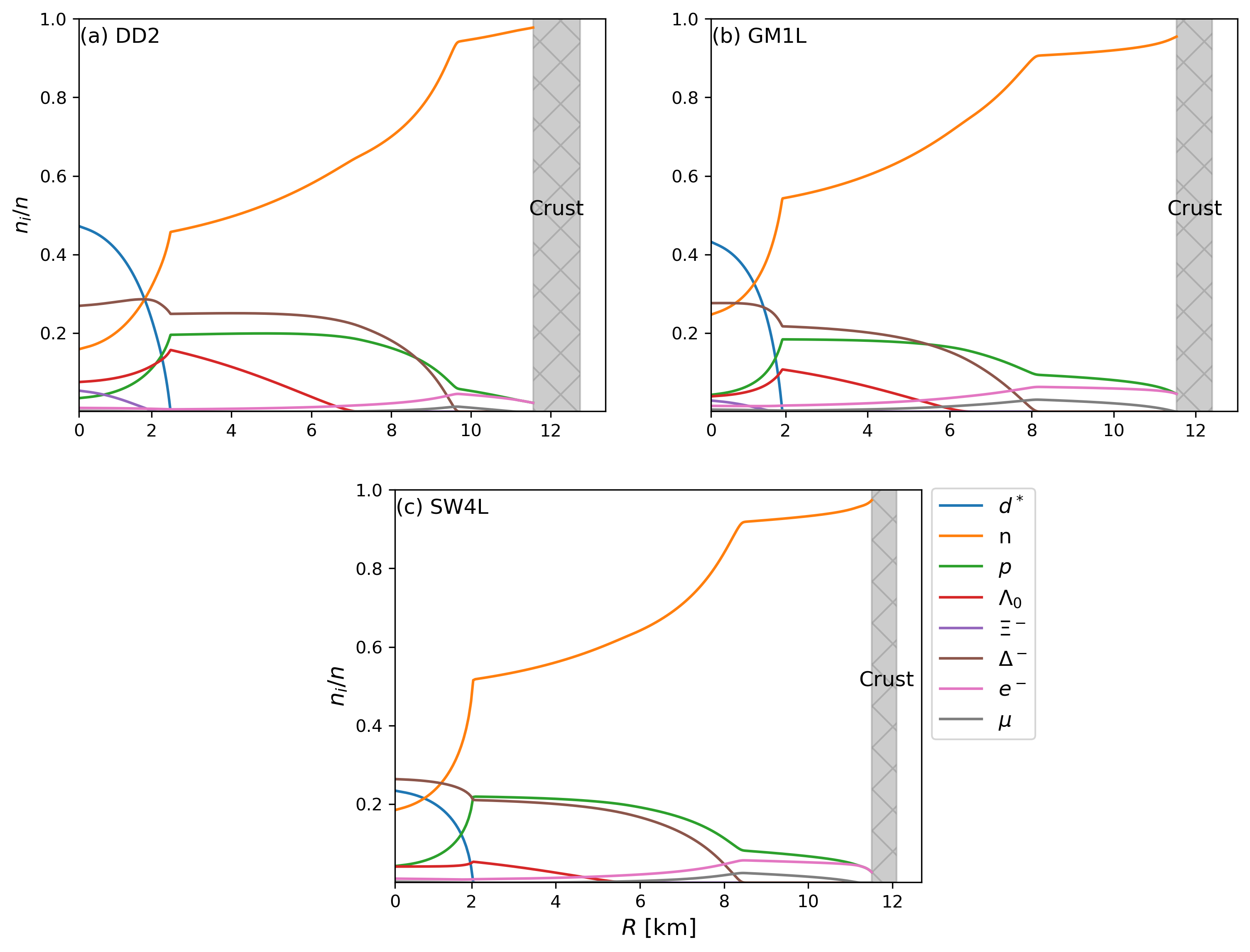}
    \caption{Particle populations of the maximum-mass stellar configurations for the DD2 (a), GM1L (b), and SW4L (c) parametrizations, considering $x_{\sigma \Delta} = x_{\omega \Delta} = 1.05$, $x_{\omega d^*} = -0.4$, and $x_{\sigma d^*} = 0.3$, with a total mass of $M = 1.95 \,M_\odot$, $M = 1.87 \,M_\odot$, and $M = 1.77 \, M_\odot$, respectively.}
    \label{fig:profs}
\end{figure*}

In Fig.~\ref{fig:profs} we show the particle population distribution in the maximum mass stellar configuration for a particular hadronic EoS characterized by coupling constant ratios \mbox{$x_{\sigma \Delta}=x_{\omega \Delta}= 1.05$}, $x_{\sigma d^*}= 0.3$, and $x_{\omega d^*}= -0.4$. Notably, the presence of the $d^*$ hexaquark becomes evident at
approximately  $n \simeq 3.60\, n_0$ in panels (a) and (c),  and at $n \simeq 3.75\, n_0$ in panel (b), significantly influencing the behaviour of all other particles in the star. Due to its high mass, the $d^*$ particle appears only deep in the cores of these NS, in the innermost part, at $R_{d^*}\sim~2$~km. 
Consequently, its impact on the NS radius is
minimal, but its presence is a key determinant for the maximum NS
mass, as demonstrated in Fig.~\ref{fig:mr}.

For the DD2 and GM1L parametrizations in panels (a) and (b) respectively, the $\Delta^-$ resonance is present at low densities, with $n \sim 1.56\, n_0$ and $n \sim 2.23\, n_0$ respectively. However, in the case of the SW4L parametrization in panel (c), the appearance of the $\Delta^-$ resonance takes place at very high densities, specifically at $n \sim 7.60, n_0$, practically at the center of the star. As a result, its presence is not as noticeable in this scenario.

In Fig.~\ref{fig:profs}, we observe another intriguing aspect of models incorporating the $d^*$ particle: due to its low mass, the $\Lambda$ emerges at lower densities compared to the $d^*$ in all three cases. However, the presence of the $d^*$ particle hinders the $\Lambda$ particles from reaching a substantial fraction in the EoS, and this holds true for all three models.

In chiral effective field theories, such as in  Ref.~\cite{Weise2020}, the absence of the $\Lambda$ particle is ascribed to higher-order 3-body forces. However, incorporation of $d^*$
particles into EoS leads to $\Lambda$ suppression mainly through two-body forces. All models featuring the $d^*$ particle exhibit similar behaviour: as soon as the $d^*$ appears, it tends to convert all other particles into itself, except for some negatively charged species that serve as compensators for its positive charge. For all models, the $\Delta^-$ particle (and to a lesser extent, the $\Xi^-$) functions as a compensator, with electrons and muons playing a smaller role in the compensation process.

The following and final results of this study consist of colored maps that establish the relationship between the $d^*$ and $\Delta$ coupling constants with various EoS and astrophysical relevant quantities. In the subsequent paragraphs, we present these results in detail. Prior to that, we would like to clarify the general behavior of these figures: on the respective parameter planes, the results are displayed as colored polygons. Although we explore the complete ranges of coupling constants as detailed earlier, there are instances of instabilities in the resulting EoS, which prevent us from obtaining results for the entire planes presented in these figures. 
Consequently, the polygons indicate the regions where we have identified an unstable  EoS behavior. The possible causes of these emerging EoS instabilities are discussed in Section~\ref{sec:summary}.

\begin{figure*}
    \centering
   \includegraphics[width=0.8\linewidth]{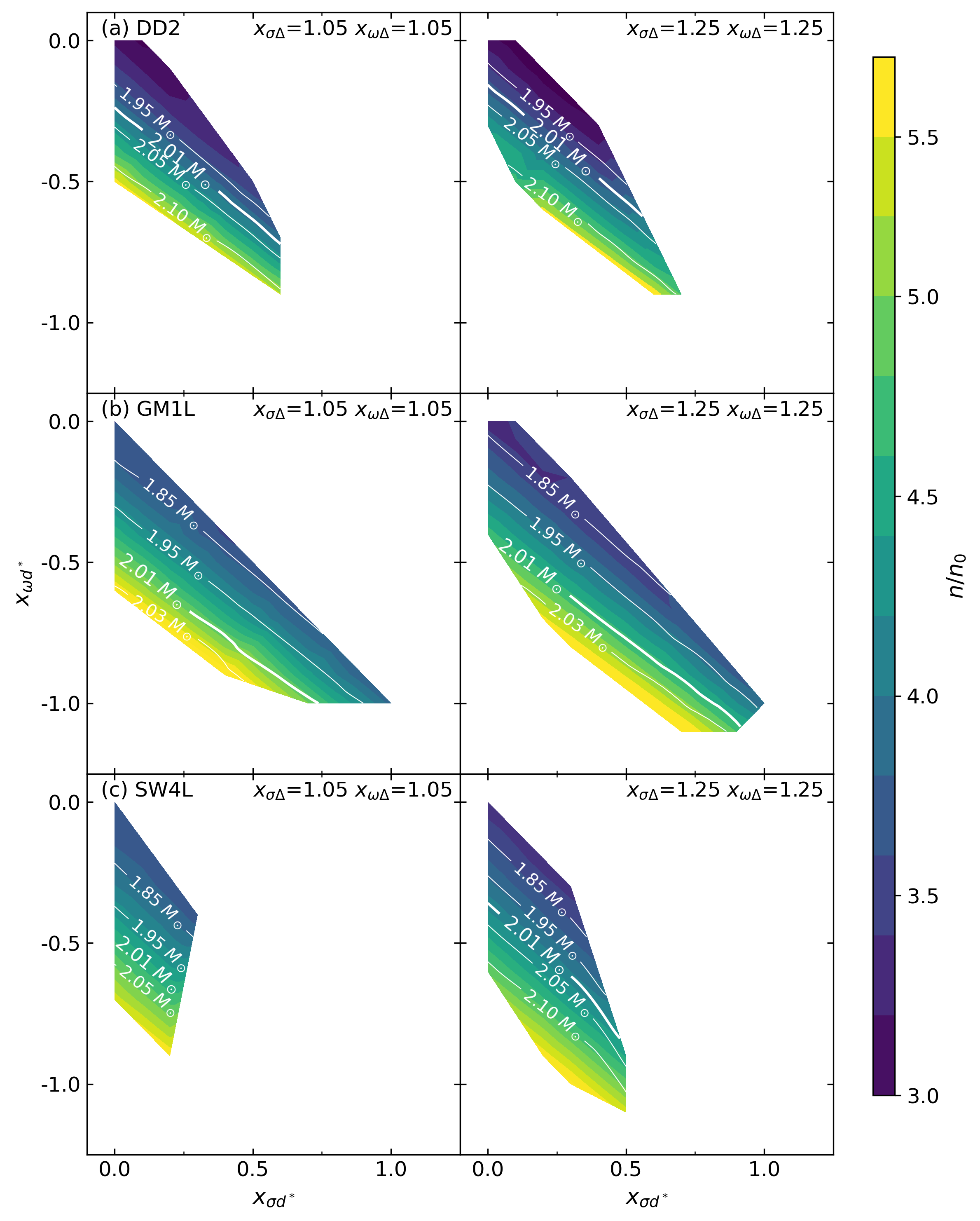}
    \caption{Onset of the dibaryon population in the $x_{\omega d^*}$--$x_{\sigma d^*}$ plane for DD2 (a), GM1L (b), and SW4L (c), considering  meson-$\Delta$ coupling constants of $x_{\sigma \Delta}$ = $x_{\omega \Delta}$ = 1.05 (left panel) and $x_{\sigma \Delta}$ = $x_{\omega \Delta}$ = 1.25 (right panel).  The color bar shows the baryon density $n$, in units of $n_0$, at which the $d^*$(2380) appears for each parametrization. The white lines show the maximum masses of NS as a function of  $x_{\omega d^*}$ and $x_{\sigma d^*}$.}
    \label{fig:onset}
\end{figure*}

Fig.~\ref{fig:onset} illustrates the color map of the \mbox{$x_{\omega d^*}-x_{\sigma d^*}$} plane for each parametrization. The left (right) panels depict the EoS with a meson-$\Delta$ coupling constant of \mbox{$x_{\sigma \Delta}=x_{\omega \Delta} = 1.05$} ($x_{ \sigma \Delta }=x_{\omega \Delta}  = 1.25$), respectively. The color bar in the figure indicates the baryonic density, 
expressed in units of the nuclear saturation density $n_0$, at which the $d^*$-hexaquark onset occurs. The white curves represent the maximum mass values achievable in the mass-radius curve of NS for the corresponding combination of EoS \mbox{($x_{\omega d^*}$, $x_{\sigma d^*}$).}

Our findings reveal that a delayed appearance of the $d^*$ particle stiffens the EoS, resulting in higher mass values in the associated curve of stellar configurations, but it also leads to the immediate destabilization of such stars. Combinations of lower values for $x_{\omega d^*}$ and $x_{\sigma d^*}$ cause the dibaryon to emerge at increasingly higher densities. The variation of the coupling constants $x_{i d^*}$ leads to a monotonic change in the $d^*$ appearance density and, consequently, contributes to a greater maximum NS mass value.

In Fig.~\ref{fig:d*_mass} we show the ratio of the gravitational stellar mass containing $d^*$ particles to the total gravitational mass of the maximum mass star, as a
function of  $x_{\sigma d^*}$ and $x_{\omega d^*}$. 
In the left (right) panels, we present the EoS for a meson-$\Delta$ coupling constant of \mbox{$x_{\sigma \Delta}=x_{\omega \Delta}= 1.05$ ($x_{ \sigma \Delta }=x_{\omega \Delta} = 1.25$)}. For the DD2 parametrization, shown in panel (a) on the left (right), the largest ratio \mbox{$M_{d^*}/M_{\rm max}=2.0 \, (1.6) \times 10^{-3}$} is obtained for $x_{\sigma d^*}=0.5$ and $x_{\omega d^*}=-0.6$ (\mbox{$x_{\sigma d^*}=0.3$} and $x_{\omega d^*}=-0.3$). For the GM1L parametrization, shown in panel (b) on the left (right), the largest ratio \mbox{$M_{d^*}/M_{\rm max}=6.3 \, (2.0) \times 10^{-3}$} is obtained for $x_{\sigma d^*}=0.9$ and $x_{\omega d^*}=-1.0$ ($x_{\sigma d^*}=0.8$ and $x_{\omega d^*}=-1.0$). 

For the SW4L parametrization, the value of $M_{d^*}/M_{\rm max}$ is $\lesssim$ $10^{-5}$ for both choices of meson-$\Delta$ coupling constants. This value is below the precision of our calculations; thus, we do not present the results in the figure. 

All parametrizations display a non-monotonic behavior with a clear maximum at certain values of the coupling constants $x_{i d^*}$. This allows us to speculate about a \textit{proper} model based on purely theoretical grounds, hypothesizing that a correct EoS should lead to the maximum possible $d^*$ content within the NS.

In Fig.~\ref{fig:d*_mass}, the alteration of the remaining baryonic coupling constants, denoted as $x_{i \Delta}$, also influences the location of the maximum $d^*$ mass peak, offering a means to adjust these constants accordingly. Additionally, as the value of a coupling constant increases and approaches the region where EoS instabilities arise, it results in a reduction of the NS mass at the point where the $d^*$ content reaches its maximum.

\begin{figure*}
    \centering
    \includegraphics[width=0.8\linewidth]{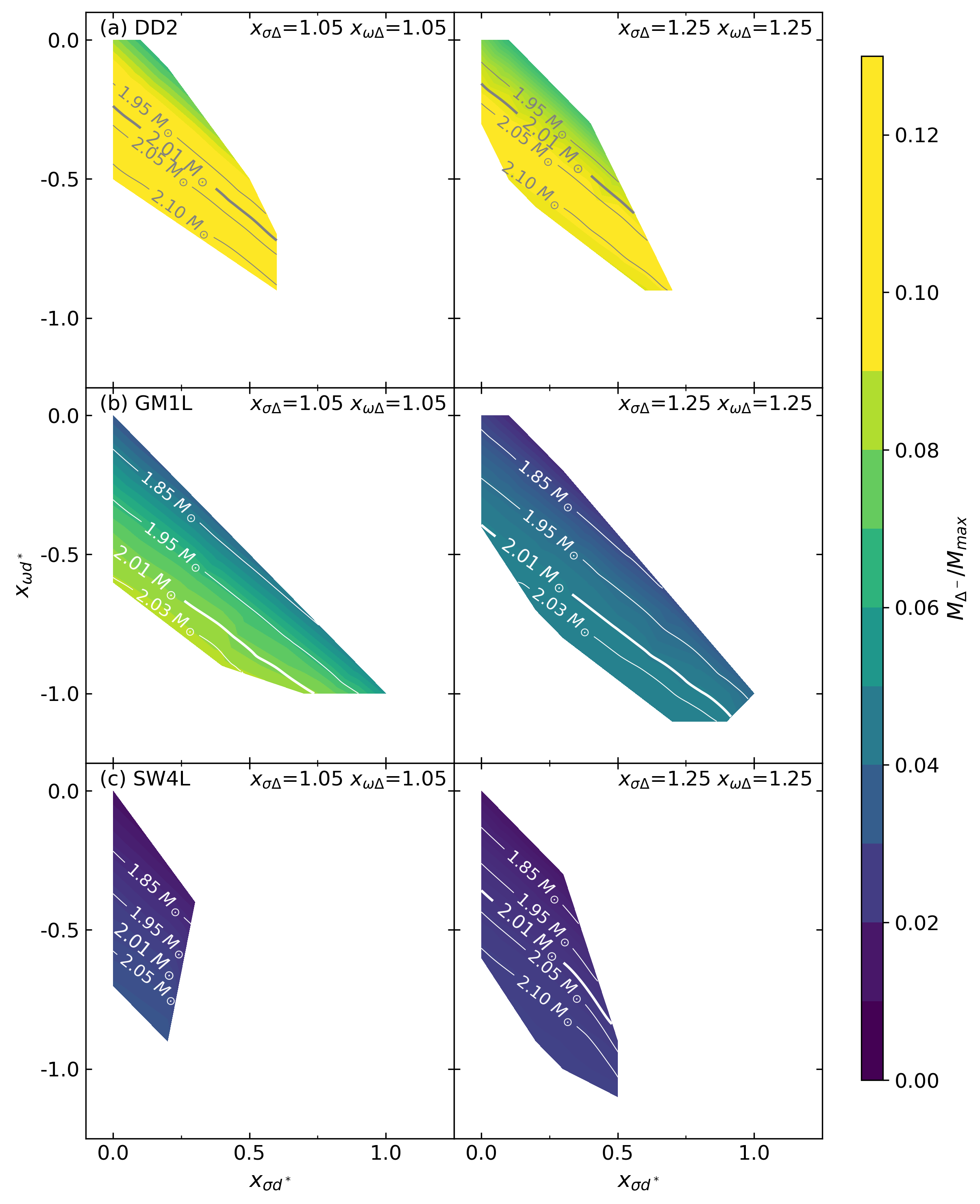}
    \caption{Ratio between the stellar mass containing $\Delta^-$ particles, $M_{\Delta^-}$, and 
    the maximum stellar mass, $M_{\rm max}$, in the $x_{\omega d^*}$--$x_{\sigma d^*}$ plane for \mbox{$x_{\sigma \Delta} = x_{\omega \Delta} = 1.05$} on the left-hand-side and $x_{\sigma \Delta} = x_{\omega \Delta} = 1.25$ on the right-hand-side). The analysis considers the DD2, GM1L, and 
    SW4L parametrizations. The color bar shows the $\Delta^-$ mass fraction in the 
    maximum-mass stellar configurations.
    The dark and white lines show the maximum masses of NS as a function of  $x_{\omega d^*}$ and $x_{\sigma d^*}$.    }
    \label{fig:delta_mass}
\end{figure*}

In Fig.~\ref{fig:delta_mass}, we present the ratio of the gravitational stellar mass containing $\Delta^-$ particles to the total gravitational mass of the maximum-mass stellar configuration, as a function of $x_{\sigma d^*}$ and $x_{\omega d^*}$. The left (right) panels depict the EoS for a meson-$\Delta$ coupling constant of \mbox{$x_{\sigma \Delta}=x_{\omega \Delta} = 1.05$} (\mbox{$x_{ \sigma \Delta }=x_{\omega \Delta}  = 1.25$}).

In the specific case of the right panel of the DD2 parametrization, the largest value of \mbox{$M_{\Delta^-}/M_{\rm max} \sim  0.1$} is achieved at the center of the colored area, where \mbox{$x_{\sigma d^*}=0.0$} and \mbox{$x_{\omega d^*}=-0.3$.} In all other panels, $M_{\Delta^-}/M_{\rm max}$ increases as $x_{\omega d^*}$ decreases for a fixed value of $x_{\sigma d^*}$. Panel (a) of 
the DD2 parametrization shows the highest \mbox{$M_{\Delta^-}/M_{\rm max}$ ratio, approximately $10\%$ of $M_{\rm max}$.} For the left panel (a), this is achieved with \mbox{$x_{\sigma d^*}=0.4$} and $x_{\omega d^*}=-1.2$. In the case of the left (right) panel (b) of the GM1L parametrization, we obtain \mbox{$M_{\Delta^-}/M_{\rm max} \sim  0.05 \, (0.09)$} by combining \mbox{$x_{\sigma d^*}=0.3$} and $x_{\omega d^*}=-0.7$ ($x_{\sigma d^*}=0.1$ and $x_{\omega d^*}=-1.0$). The lower mass fraction $M_{\Delta^-}$ is obtained for the SW4L parametrization, in panels (c), where $M_{\Delta^-}/M_{\rm max} \sim  0.03$ in the left (right) panel with $x_{\sigma d^*}=0.1$ and $x_{\omega d^*}=-1.0$ ($x_{\sigma d^*}=0.3$ and \mbox{$x_{\omega d^*}=-0.5$}).

It is interesting to note that conventional baryons, such as $\Delta$ resonances, demonstrate a monotonic behavior in $M_{\Delta^-}/M_{\rm max}$ with the variation of $x_{i d^*}$.

\section{Summary and Conclusions} \label{sec:summary}

In this study, we extended the investigation of the $d^*$-hexaquark's presence in NS, building upon the work of \citet{Mantziris:2020nsm}. To model the matter in the interior of such compact objects, we employed three different parametrizations of relativistic mean-field  models with density-dependent coupling constants, namely DD2, GM1L, and SW4L. 
The interactions among protons, neutrons, hyperons, and the $\Delta$-resonances 
are described by the exchange of $\sigma$, $\omega$, $\rho$, $\sigma^*$, and $\phi$ mesons.
For each parametrization, we incorporated the $d^*$ particle, accounting for s-wave condensation and its dispersion relation, while assuming a constant (i.e., density-independent) coupling constant associated with the dibaryon. The coupling constants for the $\Delta$-resonances were set to quasi-universal values: $x_{ \sigma \Delta }=1.05$ and $x_{\omega \Delta} =1.25$. Throughout our analysis, we explored a range of coupling values for the $d^*$ particle within the intervals $-2 \leq x_{\omega d^*}\leq 0$ and $0 \leq x_{\sigma d^*}\leq 2$. By considering different families of NS and their corresponding EoS, we constructed color maps, taking into account current observational constraints on compact stars. The main findings of our work are the following:

\begin{itemize}
\item The $d^*$ hexaquarks form a boson condensate, resulting in a softening of the
  equation of state for neutron star matter.

\item The critical density at which the $d^*$ condensate emerges lies
  between 4 and 5 times the nuclear saturation density, varying
  depending on the specific EoS. In this study, we utilized the SU(3)
  ESC08 model to determine the hyperon-meson coupling constants at
  nuclear saturation density, focusing on $x_{\sigma \Delta} = 1.05$
  and $x_{\omega \Delta} = 1.25$ values. However, it should be noted
  that values around these choices are also possible.  Since these
  coupling constants are phenomenological, they account for
  higher-order effects, such as $3-$ and $4-$body forces, and there
  are no first principle constraints to fix them at specific values.
  By varying the values of the coupling constants, one obtains EoS
  that are either stiffer or softer than the models used in our
  paper. Due to this constraint, it is not possible to reach
  definitive conclusions regarding the presence or absence of $d^*$
  hexaquarks in massive neutron stars.

\item  This indicates that $d^*$ hexaquarks exist within a sphere with a   radius of less than $\sim 2$~km in the cores of neutron stars.

\item  $d^*$ hexaquarks are found to exist only in rather massive neutron stars. 

\item Within the parameter spaces examined in our paper, the masses of
  such neutron stars cannot significantly exceed two solar masses when
  the relativistic mean-field theory is employed to model dense
  neutron star matter.

\item Further investigation is needed to determine if other theories of dense matter, possibly incorporating phase transitions to other types of matter, could lead to neutron stars where the destabilizing emergence of $d^*$ hexaquarks allows for masses exceeding two solar masses.\\

\item If this situation does not occur, one can confidently dismiss the presence of $d^*$ hexaquarks in dense neutron star matter, leading to more precise calculations of dense matter with one less degree of uncertainty.

\item The influence of the $d^*$ particle on the EoS and its
impact on the observational constraints of NS is related to the
strengthening of the attraction ($g_{\sigma d^*}$) or the repulsion
($g_{\omega d^*}$). We observed a general trend that increasing the
absolute values of the coupling constants increases the $d^*$ content
in the stellar core up to $4 \times 10^{-3}$ times the maximum stellar
mass.
\end{itemize}

In a future study, we intend to conduct minimization studies in order to investigate how the maximum $d^*$ content varies across the complete parameter space. However, due to the extensive nature of the parameter space (approximately 50 parameters), significant computational resources will be necessary for this undertaking.

\section{Appendix}
\label{sec:App}
\subsection{Meson-hyperon coupling constants}

Based on a modified SU(3) symmetry, we have used the Nijmegen extended-soft-core (ESC08) model to determine the vector meson-hyperon coupling constants. These couplings can be expressed in terms of octet-singlet coupling ratio z, the vector mixing angle $\theta_V$, and the symmetric/antisymmetric vector coupling ratio $\alpha_V$ \cite{Oertel:2016hin}. In the the SU(3) ESC08 model, $z=0.1949$, $\theta_V=37.57^{\circ}$ and $\alpha_V=1$. Therefore, for the three parametrizations considered in this work, we have \mbox{$x_{\omega \Lambda}=x_{\omega \Sigma}=0.79426$,}  and $x_{\omega \Xi}=0.588521$, where $x_{\omega Y}=g_{\omega Y}/g_{\omega N}$. Additionally, for the SW4L parametrization, $x_{\phi \Lambda}=x_{\phi \Sigma}=-0.609460$, and $x_{\phi \Xi}=-0.877583$, where $x_{\phi Y}=g_{\phi Y}/g_{\omega N}$. Note that $g_{\phi N}=x_{\phi N}~g_{\omega N}$, and $x_{\phi N}=x_{\phi \Delta}$.

\begin{table}[h]
\begin{tabular}{cccc}
\toprule
\multicolumn{1}{l}{Parametrization} & \multicolumn{1}{l}~~~{g$_{\sigma N}$}~~~& \multicolumn{1}{l}~~~{g$_{\omega N}$}~~~ & \multicolumn{1}{l}~~~{g$_{\rho N}$}~~~ \\ \midrule
DD2                                 & 10.69                              & 13.34                              & 3.627                            \\
GM1L                                & 9.572                              & 10.62                              & 8.198                            \\
SW4L                                & 9.801                              & 10.39                              & 7.818 \\ \bottomrule

\end{tabular}
\caption{Scalar ($\sigma$), vector ($\omega$), and
isovector ($\rho$) meson-nucleon coupling constants for the parametrizations used in this work.}\label{couplings}
\end{table}

\begin{table}[]
\begin{center}
\begin{tabular}{cccc}
\toprule 
~~Parameters~~ & ~~~DD2~~~ & ~~~~GM1L~~~ & ~~~~SW4L~~~\\
\midrule
$m_{\sigma}$  (GeV)    &  0.5462     &  -0.5500 & -0.5500     \\
$m_{\omega}$  (GeV)          &0.7830    &-0.7830 & -0.7826 \\
$m_{\rho}$  (GeV)          &  0.7630       & -0.7700 & -0.7753   \\
$m_{\sigma^*}$  (GeV)  &  --& --      & -0.9900         \\
$m_{\phi}$  (GeV)    &   --&--   & -1.0195         \\
$\tilde{b}_{\sigma}$         &     --             & -0.0029    &-0.0041     \\
$\tilde{c}_{\sigma}$         &      --           &   -0.0011    & -0.0038    \\
$a_{\sigma}$         &1.3576        &  --  &-- \\
$b_{\sigma}$         & 0.6344                  & -- &--          \\
$c_{\sigma}$         & 1.0054           &  --    & --    \\
$d_{\sigma}$         & 0.5758                 &  --   &--     \\
$a_{\omega}$         &1.3697           &--  &-- \\
$b_{\omega}$         & 0.4965               &--     &--     \\
$c_{\omega}$         & 0.8177                 &  --   &--     \\
$d_{\omega}$         &0.6384                 &  --  &--      \\
$a_{\rho}$         & 0.5189          & -0.3898  &-0.4703\\ \bottomrule
\end{tabular}
  \caption{Meson masses and constants of the functions of Eqs.~(\ref{eq:so_func}) and (\ref{eq:rho_func}) for the parametrizations that lead to the properties of symmetric nuclear matter at saturation density given in Table \ref{table:properties}.}
\label{table:parametrizations}
\end{center}
\end{table}

Once the vector meson-hyperon couplings are determined, the scalar meson-hyperon coupling constants, $x_{\sigma Y}$, $x_{\sigma^* Y}$, can be fitted to reproduce empirical hyperon single-particle potentials in symmetric nuclear matter at nuclear saturation density given in Ref. \cite{Spinella:2019hns}

\begin{eqnarray}
  U_{Y}^{~(N)}(n_0) = g_{\omega Y} \bar\omega + g_{\phi Y} \bar\phi
  - g_{\sigma Y} \bar\sigma,
  \label{U_nucleon-hyperon}
\end{eqnarray}
where we have considered $U_{\Lambda}^{(N)}(n_0)=-28$ MeV, $U_{\Sigma}^{(N)}(n_0)=+30$ MeV,
and $U_{\Xi}^{(N)}(n_0)=-14$ MeV. In the case of SW4L parametrization, where the strange-scalar meson $\sigma^*$ is included, we use the following potential

  \begin{equation} \label{eq:lambda-lambda-potential}
  U_{\Lambda}^{~\!(\Lambda)}(n_0) = 
	  g_{\omega \Lambda} \bar\omega
  + g_{\phi \Lambda} \bar\phi
  - g_{\sigma \Lambda} \bar\sigma
  - g_{\sigma^* \Lambda} \bar\sigma^*\,,  
\end{equation}
where $U_{\Lambda}^{(\Lambda)}(n_0)=-1$ MeV is considered to set the coupling constant
$g_{\sigma^*\Lambda}$ in
isospin-symmetric $\Lambda$-matter, $g_{\sigma^*\Sigma} =
g_{\sigma^*\Lambda}=1.924214$, $g_{\sigma^*\Xi}=7.724675$. Note that x$_{\sigma^* N}=0$, and x$_{\sigma^* N}=x_{\sigma^* \Delta}$ (see Ref. \cite{Malfatti:2020dba} for details).

\begin{table}[h]
\begin{center}
\begin{tabular}{cccc}
\toprule 
~~Parameters~~ & ~~~DD2~~~ & ~~~~GM1L~~~ & ~~~~SW4L~~~\\
\midrule
$x_{\sigma \Lambda}$     &  0.7185     &  0.7089 & 0.7606     \\
$x_{\sigma \Sigma}$            &0.5773    & 0.5030 & 0.5547 \\
$x_{\sigma \Xi}$            &  0.5276       & 0.5155 & 0.6014   \\
$x_{\omega \Lambda}$   &  & 0.7943      &          \\
$x_{\omega \Sigma}$      &   &0.7943   &         \\
$x_{\omega \Xi}$        &                 & 0.5885    &     \\
$x_{\rho \Lambda}$         &      0          &   0   & 1   \\
$x_{\rho \Sigma^{\pm}}$        &2       &  2  &1 \\
$x_{\rho \Sigma^0}$        &0       &  0  & 1 \\
$x_{\rho \Xi}$         &                  & 1   &          \\
\bottomrule
\end{tabular}
  \caption{Coupling constants ratios $x_{i H}$ ($i = \sigma, \omega, \rho$; $H = \Lambda, \Sigma, \Xi$) for the parametrizations that lead to the properties of symmetric nuclear matter at saturation density given in Table \ref{table:properties}.}
\label{table:x_ratios}
\end{center}
\end{table}

The relative isovector meson-hyperon coupling constants for DD2 and GM1L parametrizations are scaled by the
hyperon isospin as $x_{\rho Y} = 2\,\lvert I_{3Y}\rvert$. For SW4L parametrization, we have used universal isovector meson-hyperon couplings. The meson-nucleon coupling constants for the different parametrizations are listed in Table \ref{couplings}. Table \ref{table:parametrizations} contains the meson masses and the constants for the density-dependent functionals of Eqs. (\ref{eq:so_func}) and (\ref{eq:rho_func}).

\begin{table}[h]
\begin{tabular}{cc}
\toprule
\multicolumn{1}{l}{Baryon} & \multicolumn{1}{l}~~~{ Mass (MeV)}~~~ \\ \midrule
$n$                                 & 939.6                                      \\
$p$                                & 938.3                                \\
$\Lambda^0$                         &  115.7                                  \\
$\Sigma^+$                         &    1189.4                                \\
$\Sigma^0$                         &     1192.6                               \\
$\Sigma^-$                         &     1197.4                               \\
$\Xi^0$                         &         1341.9                           \\
$\Xi^-$                         &         1321.3                           \\ \bottomrule
\end{tabular}
\caption{Masses of the spin 1/2 baryon octet for the parametrizations used in this work.}\label{masses_h}
\end{table}

To improve the readability of the coupling constants, we present the ratio $x_{i H}$ ($i = \sigma, \omega, \rho$; $H = \Lambda, \Sigma, \Xi$) in Table \ref{table:x_ratios}. In addition, the masses of the spin 1/2 baryon octect for the three parametrizations are given in Table (\ref{masses_h}).

As mentioned in Section \ref{sec:RMFd*}, in addition to the hadronic decuplet, we also consider the possibility of the corresponding hyperon Fermi channel, $\Omega^-$, for which we assume the coupling constants with mesons to be universal.

\section*{Acknowledgments}

This work is supported through the U.S. National Science Foundation under Grant PHY-2012152 and the UK STFC (ST/L00478X/2, ST/V002570/1, ST/V001035/1). M.O.C, M.G.O, M.M and I.F.R-S thank CONICET and UNLP for financial support under grants PIP-0169 and 11/G187. I.F.R-S is partially supported by PICT grant 2019-0366 from ANPCyT and PIBAA grant 0724 from CONICET (Argentina). M. G. O. thanks Dr. W. Spinella for his valuable discussions and assistance regarding the SW4L parametrization.





%

\end{document}